# INSTABILITY OF A WITNESS BUNCH IN A PLASMA BUBBLE

A. Burov, V. Lebedev, and S. Nagaitsev

Fermilab, Batavia, IL 60510

*Abstract*

The stability of a trailing witness bunch, accelerated by a plasma wake accelerator (PWA) in a blow-out regime, is discussed. The instability growth rate as well as the energy spread, required for BNS damping, are obtained. A relationship between the PWA power efficiency and the BNS energy spread is derived.

## THE LU EQUATION

When the drive bunch is intense enough, it produces an electron-free plasma bubble, with an average radius, $R$, exceeding the characteristic plasma length, $R \gg k_p^{-1}$. In this case, the shape of the bubble $r_b(z)$ can be described by the Lu equation [@Lu]:

$$r_b r_b'' + 2 r_b'^2 + 1 = \frac{4\lambda(z)}{r_b^2};$$

$$\lambda(z) = \frac{N_d}{(2\pi)^{3/2} \sigma n_e} \exp\left(-\frac{z^2}{2\sigma^2}\right); \quad (1)$$

$$\int \lambda(z) dz = \frac{N_d}{2\pi n_e}.$$

Here $\lambda(z)$ is the beam line density normalized in agreement with Ref [@Lu], $N_d$ is the number of electrons in the drive bunch, $n_e$ is the plasma density and z is the distance from the drive bunch; zero boundary conditions sufficiently ahead of the bunch are assumed for $r_b$ and $r_b'$. After the substitution $r_b \to \sigma r_b$; $z \to \sigma z$, the equation is made to depend on a single parameter $q$:

$$r_b r_b'' + 2 r_b'^2 + 1 = \frac{4\lambda(z)}{r_b^2};$$

$$\lambda(z) = \frac{q}{(2\pi)^{3/2}} \exp\left(-\frac{z^2}{2}\right); \quad (2)$$

$$q = \frac{N_d}{n_e \sigma^3}.$$

Figures 1 and 2 show the bubble boundary for $q = 100$ and $q = 800$ respectfully. As one can see, the bubble is approximately spherical, except for its very frontal part, where the drive bunch is located. The bubble radius is seen to be well approximated by $R \cong \sigma q^{1/3} = (N_d / n_e)^{1/3}$, so it does not depend on the bunch length $\sigma$. Note that the related range of parameters corresponds to $R \gg \sigma$. The longitudinal electric field on the axis is given by [@Lu]

$$E_z = 2\pi n_e e r_b r_b' \approx 2\pi n_e e R \zeta = E_0 \zeta \quad (3)$$

where $\zeta = (z - z_0)/R$ is a dimensionless distance to the bubble centre, and the approximation assumes reasonable closeness to the centre. The transverse focusing field, under the same conditions, is given by [@Lu]

$$F_r = 2\pi n e r = E_0 r / R. \quad (4)$$

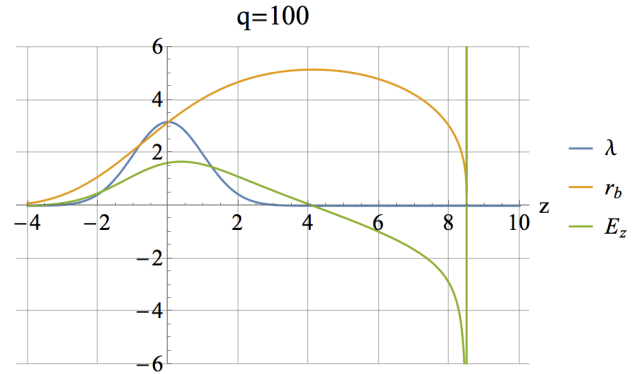

Fig. 1: Solution of the Lu equation for its parameter $q$=100. The line density $\lambda$ is in arbitrary units, the bubble radius is in the units of the rms bunch length $\sigma$ and the longitudinal field $E_z$ is in units of $4\pi n_e e \sigma$.

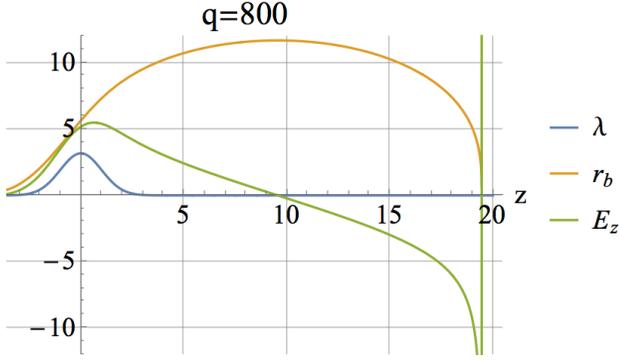

Fig. 2: Same as the Fig 1 but for $q=800$.

It is worth mentioning that, according to Eq. (2), the number of plasma ions in the bubble is proportional to the number of electrons in the drive bunch:

$$N_i = \frac{4}{3}\pi R^3 n_e \approx 4 N_d \qquad (5)$$

Let us define the transformer ratio as

$$r_t = \frac{E_0}{\langle E_z \rangle_d} \approx \frac{11}{q^{1/4}}; \quad (q \geq 100)$$

$$\langle E_z \rangle_d = \frac{\int dz \lambda E_z}{\int dz \lambda}. \qquad (5)$$

With that, the power transfer ratio can be expressed as

$$\frac{P_w}{P_d} = r_t \zeta \frac{N_w}{N_d}, \qquad (5)$$

where $N_w$ is a number of electrons in the witness bunch.

## WAKE FUNCTIONS

The Lu equation (1) allows for obtaining the longitudinal wake function for witness beam particles. Indeed, an addition of the line density perturbation

$$\delta \lambda(z) = \frac{\delta(z - z_1)}{2\pi e n_e} \qquad (6)$$

to Eq.(1) yields a step in the longitudinal field

$$\delta E_z = W_\parallel(-0) = \frac{4}{r_b^2} \qquad (7)$$

which is, by definition, the longitudinal wake function at distances $|\Delta z| \ll R, k_p^{-1}$. Note that this wake value is identical to the one for a hollow cylindrical plasma channel of the radius $r_b$, as found in Ref. [@Schroeder99]. The same value of the short-range wake is valid for dielectric channels and resistive walls as well [@BurNov, @Chao]. For any channel, the Panofsky-Wenzel relation between the longitudinal and transverse wakes is given by [@Bane]:

$$W_\perp = \frac{2}{r_b^2} \int W_\parallel dz = \frac{8 \Delta z}{r_b^4}. \qquad (8)$$

Below, we are assuming that this transverse wake is valid for the particles of a witness bunch, which length is small enough, $\Delta z \ll R, k_p^{-1}$.

## INSTABILITY

The transverse instability of an accelerated short homogeneous bunch a zero momentum spread in a plasma channel was described in Ref.[@Schroeder]; its result for a slow instability is reproduced below. Assuming that focusing scales with the energy as $k_\beta \propto \gamma^{-1/2}$, which is valid for the plasma density being kept constant, the equation of motion has been solved with the following result for the local transverse offsets along the bunch:

$$\frac{X(l,s)}{X_0} = \frac{3^{1/4}}{2^{3/2} \pi^{1/2}} \left(\frac{\gamma_0}{\gamma}\right)^{1/4} \frac{\exp(A)}{A^{1/2}} \cos\psi; \qquad (9)$$

$$\psi = \theta - A/3^{1/2} + \pi/12.$$

Asymptotically, $A \to (s/L)^{1/6}$, with the instability growth length

$$L = \frac{2^{10}}{3^9}\left(\frac{I_0}{I}\right)^2 \left(\frac{k_0 \gamma_0^{1/2} r_b^4}{8 l^2}\right)^2 \gamma';$$

$$k_0 \gamma_0^{1/2} = k_\beta \gamma^{1/2} = k_p / 2^{1/2}; \qquad (10)$$

$$I_0 = mc^3/e; \quad I = N_w ec/l.$$

For the plasma bubble, where

$$k_0^2 \gamma_0 = k_p^2/2; \quad \gamma' = k_p^2 R \zeta / 2;$$
$$r_b^2 = R^2(1 - \zeta^2); \quad R = (N_d / n_e)^{1/3}; \qquad (11)$$

the instability length is simplified to

$$L \approx 0.03 \left(\frac{N_d}{N_w}\right)^2 \frac{R^3}{l^2} \zeta(1-\zeta^2)^4. \qquad (12)$$

The growth length $L$ reaches its maximum at $\zeta = 1/3$:

$$L_{max} \approx 0.006 \left(\frac{N_d}{N_w}\right)^2 \frac{R^3}{l^2}. \qquad (13)$$

Assuming, for example, $N_d/N_w = 10$, and $R/l = 10$, we get $L_{max} \approx 60 R$. Due to the extremely slow growth of the exponent power $A$, the available acceleration length $s_{max}$ can exceed the instability length

$L_{max}$ by several orders of magnitude. Indeed, let's assume that the offset growth shown in Eq.(9) is limited by a factor of 10: $X/X_0 = 10(\gamma_0/\gamma)^{1/4}$. This limitation yields a maximum for the exponent power $A \leq A_{max} \approx 4.0$, and correspondingly,

$$s_{max} = A_{max}^6 L_{max} \approx 4 \cdot 10^3 L_{max} \approx 2 \cdot 10^5 R \quad (14)$$

for this numerical example, leading to the maximal energy

$$\gamma_{max} = A_{max}^6 L_{max} \gamma' \approx 2 \cdot 10^5 \Delta\gamma_R;$$
$$\Delta\gamma_R = \gamma' R = k_p^2 R^2 / 6. \quad (15)$$

Note that

$$\gamma_{max} \propto N_d^{10/3} N_w^{-2} l^{-2} n_e^{-1/3} \quad (16)$$

For $n_e = 10^{17} \text{cm}^{-3}$ and $N_d = 6 \cdot 10^{10}$, this results in $R \approx 100 \mu m$, $k_p R \approx 5$ and $\gamma_{max} \approx 10^6$. This limit is very sensitive to the maximally tolerable exponent power $A$, which, in its turn, depends on the accuracy of beam transfer from one plasma section into another. For example, if this exponent factor was a bit lower, $A_{max} = 3$, the maximal energy would be about 5 times lower.

In principle, a proper momentum spread in the witness bunch could suppress the instability, by means of the BNS damping mechanism [@BNS]. This option is examined in the following section.

## BNS DAMPING

In this section, the bunch is treated as consisting of two macroparticles with $N_w/2$ electrons each. The longitudinal forces acting on them follow:

$$F_1 = eE_0\zeta_1 - N_w e^2 W_\| / 4;$$
$$F_2 = eE_0\zeta_2 - N_w e^2 W_\| / 4 - N_w e^2 W_\| / 2. \quad (17)$$

Due to the difference of these two forces, the particles are getting unequal momenta, which relative difference can be presented as

$$\frac{\delta p}{p} = 2\frac{F_1 - F_2}{F_1 + F_2} = \frac{-\delta\zeta + \rho}{\zeta - \rho};$$
$$\zeta = (\zeta_1 + \zeta_2)/2; \quad \delta\zeta = \zeta_2 - \zeta_1; \quad (18)$$
$$\rho = \frac{N_w e^2 W_\|}{2eE_0} = \frac{N_w}{\pi n R^3 (1-\zeta^2)} \approx 0.3 \frac{N_w}{N_d(1-\zeta^2)}.$$

To provide BNS damping, the momentum spread has to be matched with the transverse wake [@Chao]:

$$\left.\frac{\delta p}{p}\right|_{BNS} = \frac{N_w r_0 W_\perp}{4 k_\beta^2 \gamma} = \frac{\rho\delta\zeta}{1-\zeta^2}. \quad (19)$$

From here, a condition for the distance between the macroparticles $\delta\zeta$ follows:

$$\frac{\rho - \delta\zeta}{\zeta - \rho} = \frac{\rho\delta\zeta}{1-\zeta^2}. \quad (20)$$

The relative momentum spread required by Eq.(19) can be much smaller than 100% only if the intensity parameter

$$\rho \ll 1. \quad (21)$$

With this condition, Eq.(20) yields:

$$\delta\zeta = \rho;$$
$$\left.\frac{\delta p}{p}\right|_{BNS} = \frac{\rho^2}{1-\zeta^2}. \quad (22)$$

Taking into account Eq.(18) and Eq.(5), the required momentum spread can be also presented as

$$\left.\frac{\delta p}{p}\right|_{BNS} \approx 0.1 r_t^{-2} \left(\frac{P_w}{P_d}\right)^2 \frac{1}{\zeta^2(1-\zeta^2)^3}. \quad (23)$$

Minimal value of the energy spread is achieved when the witness bunch is positioned at half-radius, $\zeta = 1/2$, which leads to a limitation:

$$\left.\frac{\delta p}{p}\right|_{BNS} \geq \frac{1}{r_t^2}\left(\frac{P_w}{P_d}\right)^2. \quad (23)$$

Thus, high power efficiency entails large energy spread.

## CONCLUSIONS

The transverse instability of the witness bunch sets a limit on the maximal energy up to which the bunch could be accelerated in the nonlinear regime of the plasma wake acceleration with a higher-intensity driving bunch. For the BNS damping, limitations on the acceptable momentum spread inevitably entail a corresponding reduction of the power efficiency of the acceleration.